\documentstyle[pre,aps,epsfig]{revtex}

\begin{document}

\draft

\wideabs{

\title{Temperature and density extrapolations in canonical
ensemble Monte Carlo simulations}

\author{A. L. Ferreira and M. A. Barroso}

\address{Universidade de Aveiro,  Departmento de  F\'\i sica,
 3810-193 Aveiro, Portugal}

\maketitle

\begin{abstract}
We show how to use the multiple histogram method to combine
canonical ensemble Monte Carlo simulations made at different
temperatures and densities. The method can be applied to study
systems of particles with arbitrary interaction potential and to
compute the thermodynamic properties over a range of temperatures
and densities. The calculation of the Helmholtz free energy
relative to some thermodynamic reference state  enables us to
study  phase coexistence properties. We test the method on the
Lennard-Jones fluids for which many results are available.
\end{abstract}


}
%end wideabs

\section{Introduction}
Histogram and multiple-histogram methods have been proposed as an
optimized way of analyzing Monte Carlo data\cite{FS88,FS89,S93}.
These methods can be included in the more general class of
reweighting methods\cite{marinari96}. The idea is to combine a
given set of standard Monte Carlo simulations to get improved
estimates of observables in a given parameter region. A related
idea is to sample a suitably chosen probability distribution
rather than a given statistical mechanics ensemble. The sampling
distribution is such that the configuration space visited  is
typical of the interval of thermodynamic parameters of interest
thus allowing the reconstruction of the appropriate statistical
mechanics ensemble. The methods of umbrella sampling
\cite{torrie77,valleau91,brilliantov98},
multicanonical\cite{berg92,janke94} and expanded ensemble
methods\cite{lyubartsev92,escobedo96}  can be seen as belonging to
this class.

In this article we show how to combine NVT Monte Carlo simulations
made at different temperatures and volumes. Our method is a
generalization to volume extrapolations of the multiple histogram
method. We further show that the method can be applied not only to
systems of particles that interact through interaction potentials
that have a simple scaling with particle distance but also to
those with arbitrary distance dependence. Furthermore, as we are
able to calculate relative free energies as a function of volume
and temperature, the method can be applied to study phase
coexistence properties\cite{valleau91,brilliantov98}.

Several simulation methods have been proposed to study phase
coexistence properties. In the Gibbs ensemble Monte
Carlo\cite{pana87} two simulation boxes equilibrate by exchanging
particles and volume and the system separates into two phases,
each one located in one of the boxes. Grand-Canonical ensemble
simulations with multiple histogramming have been used to study
critical properties and finite-size scaling in fluid
systems\cite{caillol98,wilding95,bruce92}. However, due to the low
probability of particle exchanges or insertions at high densities
these methods cannot be used to study dense phases. For such
systems special methods have been suggested that rely on the
calculation of the absolute free energies of the two
phases\cite{meijer90}. Recently a new method based on Gibbs-Duhem
integration was proposed and the concept of pseudo-ensembles was
introduced\cite{kofke93,mehta95,escobedo98}. For all of these
methods the use of the multiple histogram technique can be a
valuable auxiliary tool \cite{meijer90,escobedo98}. Our method can
also be applied to the study of solid fluid
coexistence\cite{ferreira99}.

\section{The Method}
\label{sample}

Consider a system of N interacting particles contained in a box of
volume $V_0$. For simplicity we consider a pairwise additive
interaction potential and $\{\vec r_i\}$ denotes a a given
configuration of particle coordinates. The total potential energy
of the system in a configuration is  given by $E(\{\vec
r_i\})=\sum_{<i,j>} u(\vert \vec r_i -\vec r_j\vert)$, where the
sum runs over all pairs of particles. Uniformly expanding the
system from the volume $V_0$ to the volume $V$ changes the
configuration from $\{\vec r_i\}$ to $\{\vec r_i\prime\}$, such
that $\vec r_i\prime=(V/V_0)^{1/3} \vec r_i$. The energy of the
system of volume $V$ in the new configuration is given by,
$E(\{\vec r_i\prime\})=\sum_{<i,j>} u((V/V_0)^{1/3}\vert \vec r_i
-\vec r_j\vert)$.

We will show  that it is always possible to find a set of $n_c$
variables, $C_n(\{\vec r_i\})$, with $0\le n \le n_c-1$,  that
depend on the particle coordinates. These variables can be seen as
coordinates of a column vector, $\vec C=(C_0,C_1,...,C_{n_c-1})$.
Their choice is arbitrary provided two properties are fulfilled.
First, it should be possible to write the potential energy  in
terms of these variables. Second, there is a known linear relation
between the  value of the variables in the expanded system of
volume V and their value for the system of volume $V_0$:
\begin{equation}
\vec C(\{\vec r_i\prime\})={\bf M} \cdot \vec C(\{\vec r_i\}),
\label{eqn:2}
\end{equation}
being ${\bf M}$ a square matrix with coefficients that depend
only on $V$ and $V_0$.

For example, for the Lennard-Jones potential, $u(r)=4\epsilon
[(\sigma/r)^{12}-(\sigma/r)^6]$, the vector $\vec C$ can be chosen
with only two components, $C_0$ and $C_1$
\begin{mathletters}
\label{eqn:3}
\begin{eqnarray}
C_0(\{\vec r_i\})&=&\sum_{<i,j>}
\left(\frac{\sigma}{r_{ij}}\right)^{12}, \\ C_1(\{\vec
r_i\})&=&\sum_{<i,j>} \left(\frac{\sigma}{r_{ij}}\right)^6,
\end{eqnarray}
\end{mathletters}
satisfying the two properties mentioned above.

For an arbitrary potential it may not be possible  to find
variables $C_n$ with a given volume scaling as
in the Lennard-Jones case. However, there is always a method
based on volume expansions that we describe next.
We define the coefficient $C_n$ from the volume derivatives of
$E((V/V_0)^{1/3}\{\vec r_i\})$:
\begin{equation}
C_n(\{\vec r_i\})=\left(\frac{\partial^n E((V/V_0)^{1/3}\{\vec r_i\})}
{\partial V^n}\right)_{V_0}.
\label{eqn:4}
\end{equation}
The two properties are fulfilled since the energy of a given configuration is
$E(\{\vec r_i\})=C_0(\{\vec r_i\})$
and the series expansion,
\begin{equation}
C_n(\{\vec r_i\prime\})=\sum_{l=n}^\infty \frac{C_l(\{\vec
r_i\})}{(l-n)!} (V-V_0)^{l-n}, \label{eqn:5}
\end{equation}
provide the linear relation (\ref{eqn:2}). However the vector
$\vec C$ has an infinite number of components. In practical
numerical work the above expansion needs to be stopped at a
sufficiently high-order. As it will be seen below the
approximation introduced can be controlled either by increasing
the order of the approximation or by combining simulations at
closer densities.

We denote the density of states with variables $\vec C(\{\vec
r_i\prime\})$ in some neighborhood of $\vec c(V)$ for a system of
volume V, by $\Omega(\vec c(V),V)$. This quantity can be obtained
from a phase space integration,

\begin{equation}
\Omega(\vec c(V),V)=\int_V d\vec r_1\prime \dots d\vec r_N\prime \
\ \delta(\vec C(\{\vec r_i\prime\})-\vec c(V)). \label{eqn:6a}
\end{equation}

Changing the variables of integration, $\vec r_i\prime =
(V/V_0)^{1/3} \vec r_i$ and noting relation (\ref{eqn:2}) between
the vector $\vec C(\{\vec r_i\prime\})$ and $C(\{\vec r_i\})$ we
can write,
\begin{eqnarray}
\Omega(\vec c&&(V),V)= \nonumber \\ && \left( \frac{V}{V_0}
\right)^N \int_{V_0} d\vec r_1 \dots d\vec r_N \ \ \delta( {\bf M}
(\vec C(\{\vec r_i\})-\vec c(V_0)) ), \label{eqn:6b}
\end{eqnarray}
where
\begin{equation}
\vec c(V)={\bf M} \cdot \vec c(V_0).
\label{eqn:7a}
\end{equation}
Using the property of the Dirac delta function, $\delta({\bf M}
(\vec C(\{\vec r_i\})-\vec c(V))) = \delta(\vec C(\{\vec
r_i\})-\vec c(V)) /\vert {\bf M} \vert$, where $\vert {\bf M}
\vert$ is the determinant of the matrix ${\bf M}$, we see that the
densities of states at different volumes are related by

\begin{equation}
\Omega(\vec c(V),V)\  d\vec c(V) =\bigl (\frac{V}{V_0}\bigr )^N
 \Omega(\vec c(V_0), V_0)\ d\vec c(V_0),
\label{eqn:7}
\end{equation}
with $\vec c(V)$ and $\vec c(V_0)$  related by equation (\ref{eqn:7a}).

Suppose that we perform several Monte Carlo simulations at inverse
temperature $\beta_i$ and volume $V_i$, $1\le i\le R$. Each simulation provides
an estimate of the density of states:

\begin{equation}
\Omega(\vec c(V_i),V_i)\  \delta \vec c(V_i) \approx
e^{\beta_i (E(\vec c(V_i))-f_i)} \frac{h_i(\vec c(V_i))}{M_i},
\label{eqn:8}
\end{equation}
where $E(\vec c(V_i))$ is the potential energy of the system,
$h_i(\vec c(V_i))$ is the histogram of the variables $C_n$
measured in the simulation $i$, $\delta \vec c(V_i)$ is the
histogram bin size, $M_i$ is the number of measures and $f_i$ is
the Helmholtz free energy at inverse temperature $\beta_i$ and
volume $V_i$. The values of $f_i$ are not known by now but will be
self-consistently determined later.

We use (\ref{eqn:7}) to relate the density of states at volume
$V$ to the density of states estimated at the simulation volume
by the above equation,
\begin{equation}
\Omega(\vec c(V),V)\; \delta \vec c(V) \approx
\left(\frac{V}{V_i}\right)^N e^{\beta_i (E(\vec c(V_i))-f_i)
}\frac{h_i(\vec c(V_i))}{M_i}. \label{eqn:9}
\end{equation}
The estimates of the density of states given by each
of the $R$ simulations are now combined\cite{FS89},
\begin{eqnarray}
\Omega(\vec c(V),&&V)  \; \delta \vec c(V) = \nonumber \\ &&
\sum_{i=1,R} p_i\ \left(\frac{V}{V_i}\right)^N e^{\beta_i (E(\vec
c(V_i))-f_i)} \frac{h_i(\vec c(V_i))}{M_i}, \label{eqn:10}
\end{eqnarray}
assigning to each of them a weight $p_i$. The normalized
($\sum_{i=1}^R p_i=1$)  weights  are obtained from the condition
of minimization of the statistical uncertainty on the density of
states,
\begin{equation}
\delta^2 \Omega(\vec c(V),V)= \overline{\Omega^2(\vec c(V),V)}
-\overline{\Omega(\vec c(V),V)}^2. \label{eqn:11}
\end{equation}
The number of measures in each bin of the histogram
is a random variable. Neglecting the correlations between the measures
and using the independence of the simulations we have,
\begin{equation}
\overline{h_i(\vec c(V_i))h_j(\vec c(V_j))} - \overline{h_i(\vec c(V_i))}
\ \overline{h_j(\vec c(V_j))}  \approx
\overline{h_i(\vec c(V_i))} \delta_{i,j}.
\label{eqn:12}
\end{equation}

The result for the weights is:
\begin{eqnarray}
p^{-1}_i &=& e^{\;\beta_i ( E(\vec c(V_i)) -f_i)} \nonumber \\ & &
\times \sum_{l=1}^R \left(\frac{M_l}{M_i}\right)
\left(\frac{V_l}{V_i}\right)^N e^{-\beta_l (E(\vec c(V_l))-f_l)}.
\label{eqn:13}
\end{eqnarray}
The partition function at inverse temperature $\beta$ and volume $V$
is thus:
\begin{equation}
Z(\beta,V)=\sum_{\vec c(V)} \sum_{i=1}^R \frac{h_i(\vec c(V_i)) \
 e^{-\beta E(\vec c(V))} }{\sum_{l=1}^R M_l (V_l/V)^N \
e^{-\beta_l (E(\vec c(V_l))-f_l)}},
\label{eqn:14}
\end{equation}
and the canonical average of any function $f(\vec c(V))$ is

\begin{eqnarray}
\langle f \rangle &=&\frac{1}{Z(\beta,V)} \nonumber \\ & & \times
\sum_{\vec c(V)} \sum_{i=1}^R \ \frac{f(\vec c(V))\ h_i(\vec
c(V_i)) \
 e^{-\beta E(\vec c(V))} }{\sum_{l=1}^R M_l (V_l/V)^N
e^{-\beta_l (E(\vec c(V_l))-f_l)}},
\label{eqn:15}
\end{eqnarray}
where $\sum_{\vec c(V)}$ is a sum over bins in the multidimensional $\vec c$
space.

For  the expansion (\ref{eqn:5}), the system pressure, $P(\beta,V)$,
is obtained directly from $C_1$:

\begin{equation}
P(\beta,V)= \frac{N}{\beta V} - \langle C_1 \rangle
\label{eqn:16}
\end{equation}
It is clear that in the actual calculations there is no need to compute
the histograms.
Denoting by $\vec c^{i,j}$ the measure j ($ 1\le j \le M_i $)
in the simulation number i we have:

\begin{eqnarray}
\langle f\rangle &=& \frac{1}{Z(\beta,V)} \nonumber \\
&\;\;\:\:\times& \! \sum_{i=1}^R \! \sum_{j=1}^{M_i} \frac{f(\vec
c^{i,j}(V)) e^{-\beta E(\vec c^{i,j}(V))} }{ \sum_{l=1}^R M_l
(V_l/V)^N e^{-\beta_l (E(\vec c^{i,j}(V_l))-f_l)}},
\end{eqnarray}
where
\begin{eqnarray}
Z(&&\beta,V) = \nonumber \\ & & \sum_{i=1}^R \sum_{j=1}^{M_i} \
\frac{ e^{-\beta E(\vec c^{i,j}(V))} }{\ \sum_{l=1}^R M_l
(V_l/V)^N e^{-\beta_l (E(\vec c^{i,j}(V_l))-f_l)}}.
\end{eqnarray}

One should remark that in the above expression the values $\vec
c^{i,j}(V_l)$, with $l=i$  are measured while the corresponding
coefficients with $l\neq i$ as well as $\vec c^{i,j}(V)$ are
computed from the measured values using equation (\ref{eqn:7a}).
The free energies $f_i$ are self-consistently obtained from the
conditions $f_i=-\beta_i^{-1} \ln Z(\beta_i,V_i)$ and by setting
$f_1=0$. Thus, we are able to compute free energies relative to
some thermodynamic state $\beta_1$, $V_1$.

\section{Application to the Lennard-Jones Fluid}

We measure values of $\vec C$ every 10~MCS/\textit{N} and the
simulation lengths were $10^5$~MCS/\textit{N}. The value of the
cut-off radius was always equal to half the side of the simulation
box. Standard long range corrections were added to the measured
values at the end of the simulation.

We first considered the choice
(\ref{eqn:3}) for the vector $\vec C$.
Two sets of simulations, with 108 particles,
were done at two reduced temperatures, $T^*_1= 1.15$  ,
and $T^*_2=1.3$. For the first temperature
we made 40 simulations at equally spaced reduced densities:
$0.02\le \rho^* \le 0.8$. For the second temperature we made
simulations at densities $\rho^*_i=0.02*1.1^{i-1}, \ 1\le i\le 40$.

\begin{figure}
\begin{center}
\psfig{figure=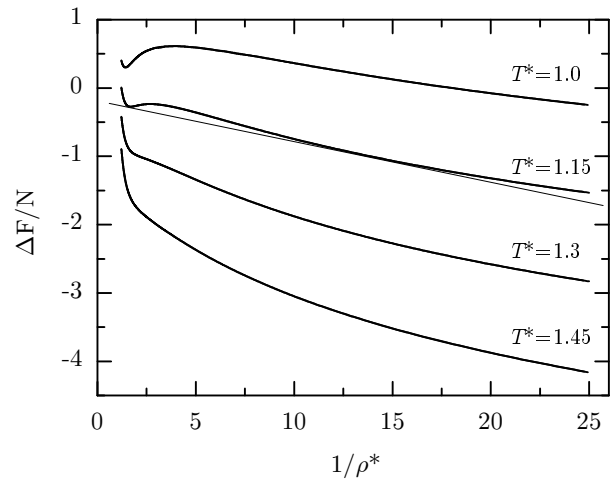}
\end{center}
\caption{ Relative Helmholtz free energy as a function of volume per particle
at four different temperatures $T^*=1.0, 1.15, 1.3$ and $1.45$. The results
obtained from the two simulation sets at temperatures  $T_1^*=1.15$ and
$T_2^*=1.3$ are plotted and they are not distinguishable. We also show the
double tangent straight line for $T*=1.15$.
\label{fig:1}}
\end{figure}

Every pair of simulations close in density  were combined using
the proposed method to obtain results for densities in between the
two simulations and for a given range of temperatures (above and
below the simulation temperature). From the free energy values
obtained we built the volume and temperature dependence of the
free energy. In Fig. \ref{fig:1} we show the free energy as a
function of volume per particle at four different temperatures
$T^*=1.0, 1.15, 1.3$ and $1.45$. In this figure we also compare
the extrapolations obtained from each of the two sets of
simulations made at $T_1^*=1.15$ and $T_2^*=1.3$. The curves from
these two simulations are nearly coincident and they are not
distinguishable in the figure. For $T^*=1.15$ we also show the
common tangent straight line at the liquid and gas coexisting
phases. The double tangent construction allows us to find the
volumes of the coexisting phases at each temperature. A new  set
of simulations with 256 particles at $T^*=1.3$ and densities
$\rho^*_i=0.1*1.047^{i-1}, \ 1\le i\le 40$ was also done.

\begin{figure}
\begin{center}
\psfig{figure=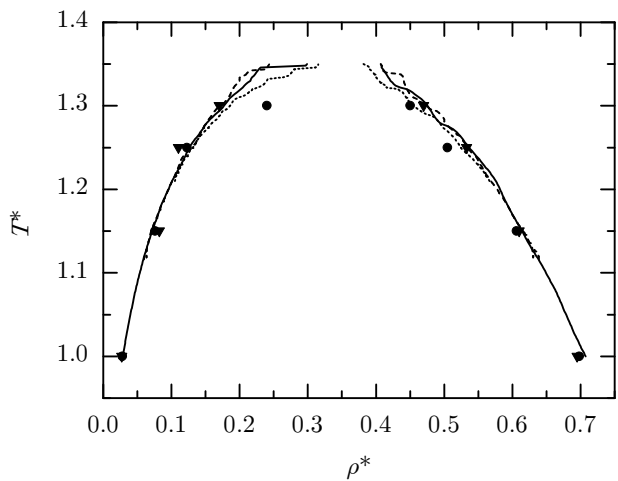}
\end{center}
\caption{
Liquid-Gas phase diagram of the three dimensional Lennard-Jones model.
 Solid line: Simulation temperature $T^*=1.3$ and 108 particles;
Dashed line: $T^*=1.15$ and $108$ particles;
Dotted line: $T^*=1.3$ and $256$ particles;
Circles and Triangles are Gibbs Ensemble
results from reference \protect\cite{pana87}
with 500 particles and 300 particles
respectively.
\label{fig:2}}
\end{figure}

In Fig \ref{fig:2} we show the phase
diagram computed from each set of simulations.

\begin{figure}
\begin{center}
\psfig{figure=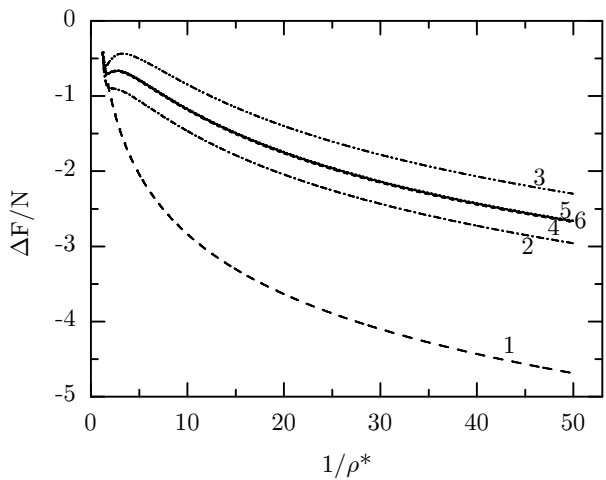
}
\end{center}
\caption{Convergence of the free energy with an increasing number
of coefficients in the expansion based in Eq. (\ref{eqn:5}).
Results with 1 to 6 coefficients correspond respectively to
long-dashed, dot dashed, dot-dot-dashed, short-dashed, dotted and
solid curves. The curves obtained with 4, 5 and 6 coefficients are
nearly coincident. \label{fig:3}}
\end{figure}

In order to ascertain the usefulness of volume expansion method
(\ref{eqn:5}) we also made simulations where the variables, $C_n,
\ \  0\le n\le 5$, were measured.  This set of simulations was
done for a system of 108 particles at $T^*=1.3$ and at the same
densities chosen before. In Fig \ref{fig:3} we show the
convergence of the results for the relative free energy as a
function of volume for a temperature $T^*=1.15$ as we include an
increasing number of coefficients in the expansion (\ref{eqn:5}).
The curves obtained with 4, 5 and 6 coefficients are nearly
coincident and agree with results obtained from the choice based
on equation (\ref{eqn:3}).

\section{Conclusions}
\label{concs} We have proposed a method which allows simultaneous
extrapolations in volume and temperature based on the multiple
histogram method. An arbitrary number of Monte Carlo simulations
made in the canonical ensemble can be combined providing improved
estimates of thermodynamic properties. We show test bed results on
the three-dimensional Lennard-Jones system which confirm that the
method works well and that the volume expansion scheme based on
equation (\ref{eqn:5}) can be used with a good control of the
approximations involved. Calculation of relative Helmholtz free
energy coupled with the  double tangent construction allows an
efficient determination of the phase diagram. The method is
general and can be applied to interaction potentials that do not
have a simple scaling with system volume.

\section*{Acknowledgements}
We thank Professors E. J. S. Lage, S. K. Mendiratta, T. P. Gasche
and J. M. Pacheco for a critical reading of the manuscript. A. L.
Ferreira gratefully acknowledges J. M. Pacheco and J. P. Ramalho
for fruitful discussions. This work was partially supported by the
projects PRAXIS/2/2.1/299/94 and PRAXIS/2/2.2/FIS/302/94.

\end{document}